\documentclass[useAMS,usenatbib]{mn2e}
\usepackage{graphicx}
\def\new#1{{#1}}

\title[A state transition of GX 339--4]
{INTEGRAL/RossiXTE high-energy observation of a state transition of GX 339--4}
\author[T. Belloni et al.]{T. Belloni$^{1}$
\thanks{E-mail:belloni@merate.mi.astro.it}, 
I. Parolin$^{1}$, 
M. Del Santo$^{2}$, 
J. Homan$^{3}$,
P. Casella$^{1}$,
R.P. Fender$^{4}$,\and
W.H.G. Lewin$^{3}$,
M. M\'endez$^{5}$,
J.M. Miller$^{6}$,
M. van der Klis$^{7}$\\
$^{1}$INAF-Osservatorio Astronomico di Brera, Via E. Bianchi 46, I-23807 Merate
(LC), Italy\\
$^{2}$INAF-Istituto di Astrofisica Spaziale e Fisica Cosmica di Roma, 
	Via Fosso del Cavaliere 100, I-00133 Roma, Italy\\
$^{3}$Center for Space Research, Massachusetts Institute of Technology, 
	77 Massachusetts Avenue, Cambridge, MA 02139-4307, USA\\
$^{4}$School of Physics and Astronomy, University of Southampton, Southampton, Hampshire SO17 1BJ\\
$^{5}$SRON, Netherlands Institute for Space Research, Sorbonnelaan 2, 3584 CA Utrecht, The Netherlands\\
$^{6}$Center for Space Research, Massachusetts Institute of Technology, 77 Massachusetts Avenue, Cambridge, MA 02139-4307, USA\\
$^{7}$Astronomical Institute ``Anton Pannekoek'', University of Amsterdam, and Center for High Energy Astrophysics, Kruislaan 403,\\ 1098 SJ, Amsterdam, The Netherlands\\
}

\begin{document}

\date{
Accepted 2005 December 29 Received 2005 November 10}

\pagerange{\pageref{firstpage}--\pageref{lastpage}} \pubyear{2006}

\maketitle

\label{firstpage}

\begin{abstract}
On 2004 August 15, we observed a fast (shorter than 10 hours) state transition in the bright black-hole 
transient GX 339--4 simultaneously with RossiXTE and INTEGRAL. 
This transition was evident both in timing and spectral properties. 
Combining the data from PCA, HEXTE and IBIS, we obtained good
quality broad-band (3-200 keV) energy spectra before and after the transition. These spectra
indicate that the hard component steepened.  Also, the  high-energy 
cutoff that was present at $\sim$70 keV before the transition was not detected after the transition.
This is the first time that an accurate determination of the broad-band spectrum across such a transition has been measured  on a short time scale. It 
shows that, although some spectral parameters do not change abruptly through the transition, the
high-energy cutoff increases/disappears
rather fast. These results constitute a benchmark on which to 
test theoretical models for the production of the hard component in these systems.

\end{abstract}

\begin{keywords}
X-ray: binaries -- accretion: accretion discs -- black hole: physics
\end{keywords}

\section{Introduction}

Black-hole candidate X-ray binaries (BHCs) are known to show transitions between
different spectral states since the first observations of Cyg X-1
(Tananbaum et al. 1971). When X-ray instrumentation became sufficiently
sophisticated to allow detailed variability studies on short time scales, the
definitions of states were refined to include fast timing properties
(van der Klis 1995, 2005; McClintock \& Remillard 2005). 
The number and defining properties of these
states have changed with time (see e.g. Homan et al. 2001), but it is now
clear that fast variability is key ingredient which needs to be
considered in order to have a complete view of the states and state
transitions.
The spectral evolution of BHCs has recently been described in terms of the
pattern in an X-ray hardness-intensity diagram (HID) (see Homan \& Belloni
2005; Belloni et al. 2005; Belloni 2005).
Original states are found to correspond to
different branches/areas of a $q$-like HID pattern. 
Four main states are identified within this framework. Two of them correspond
to the original states discovered in the seventies.
The Low/Hard State (LS), observed usually at the beginning and at the end
of an outburst, and the most common state for the persistent system Cyg X-1 (see e.g.
Dove et al. 1998; Pottschmidt et al. 2003; Cadolle Bel et al. 2005),
is identified by the presence of strong ($> 30$\% fractional rms) band-limited
noise in the power spectrum and by a hard energy spectrum. The High/Soft 
State (HS) shows weak variability, in the form of a few \% fractional rms power-law
component in the power spectrum, and an energy spectrum dominated by
a thermal disc component, with the presence of an additional weak power-law
component. The HS is usually observed, if at all, in the central intervals of an 
outburst; it is the most common state in the persistent systems LMC X-1 and
LMC X-3 (Nowak et al. 2001; Wilms et al. 2001; Haardt et al 2001). 
For a comparative example of these two states, see Belloni et al. (1999).
In between these two well-established states, the situation is rather complex and
has led to a number of different classifications. Homan \& Belloni (2005)
identify two additional states, clearly defined by spectral/timing transitions.
In the evolution of a transient, after the LS comes a transition to the Hard
Intermediate State (HIMS): the energy spectrum softens as the combined result
of a  steepening of the power-law component and the appearance of a thermal
disc component. At the same time, the characteristic frequencies in the power
spectrum increase and the total fractional rms decreases. A type-C QPO (see
Casella, Belloni \& Stella 2005) appears, also with centroid frequency increasing as
the source softens (see Belloni et al. 2005). The transition to the Soft-Intermediate
State is very sharp (sometimes over a few seconds, see Nespoli et al. 2003)
and is marked by the disappearance of the type-C QPO and by the appearance 
of a type-B QPO. In the 3-20 keV range, the corresponding variations in the
energy spectrum are rather minor (Belloni et al. 2005). Together with the association
of this transition with the ejection of fast relativistic jets, this has led to the
identification of a ``jet line'' in the HID, separating these states (Fender, Belloni \&
Gallo 2004). The jet line can be crossed more than once during an outburst (as in the 
case of XTE J1859+226:  Casella et al. 2004; Brocksopp et al. 2002).

While the physical nature of the soft component is commonly associated with an
optically thick accretion disc, there is no consensus as to the origin of the
hard power-law component, where a power law
is probably a simplification of a much more complex
reality. The energy spectra include also additional components,
important for the physics of accretion around black holes, as
emission line features  (see e.g.
Reynolds \& Nowak 2003, Miller et. al 2002, Miller et al., 2004a)
and Compton reflection bumps (see e.g. George \& Fabian 1991,
Zdziarski et al. 2001, Frontera et al. 2001). 
The spectral slope of the hard component is around $\sim$1.6 for the LS, 
1.6-2.5 in the SIMS/HIMS and 2.5-4 for the HS.
One important parameter to measure is the presence/absence of a high-energy
cutoff in the spectrum, for which observations at energies $>$20 keV are 
necessary. It is known since a long time (Sunyaev \& Tr\" umper 1979)
that the LS spectrum does show such a cutoff around $\sim$100 keV. A 
comparative measurement of a number of systems with CGRO/OSSE has
been presented by Grove et al. (1998). Here the energy spectra can clearly
be divided into two classes: the ones with a strong soft thermal component
and {\it no} evidence of a high-energy cutoff until $\sim$1 MeV, and those
with no soft component and a $\sim$100 keV cutoff. 
While the second class can 
clearly be identified with the LS, an identification of the first class is not clear.
Since both the SIMS/HIMS and the HS show a strong soft component (see
e.g. Nespoli et al. 2003) and OSSE integration times are very long, different states
could be mixed, 
making a precise identification of the source state uncertain. 
More recently, Zdziarski et al. (2001) and Rodr\' \i guez et al. (2004) measured the
high-energy spectrum of GRS 1915+105 and found no direct evidence for a 
high-energy cutoff, but spectra which appeared to contain two components.
A hybrid model was found to be necessary to interpret the spectra of Cyg X-1
(Malzac et al., 2005; Cadolle Bel et al. 2005).

The nature of the hard component in BHCs is not known, nor is it known
whether the ones observed in the HS and the LS have the same physical 
origin. For the LS, 
different models include thermal comptonization (e.g. Dove et al. 1997;1998),
bulk motion comptonization (e.g. Laurent \& Titarchuk 1999), jet syncrotron plus comptonization
(e.g. Markoff, Falcke \& Fender 2001; Markoff, Nowak \& Wilms 2005).
For the HS, the situation is less clear, partly because of the limited statistics available in the
soft spectra and partly because of the absence of an observed characteristic energy such
as a high-energy cutoff. Possible models involve non-thermal/hybrid  comptonization (e.g. Gierli\'nski et al. 1999; Zdziarski et al. 2001; Malzac et al. 2005; Cadolle Bel et al. 2005) and 
bulk-motion comptonization  (e.g. Turolla, Zane \& Titarchuk 2002).

GX 339--4 was
one of the first two BHCs for which a complete set of transitions has been observed and studied
(see Miyamoto et al. 1991; Belloni et al. 1997; M\'endez \& van der Klis 1997).
This transient black-hole candidate is known to spend long
periods in outburst. Historically, it was found prevalently in a
hard state, although several transitions were reported (Maejima et al. 1984;
Ilovaisky et al. 1986; Miyamoto et al. 1991).
From the launch of the Rossi X-Ray Timing
Explorer (RXTE) until 1999 it remained detectable with the RXTE All-Sky Monitor, mostly in the
Low/Hard state but with a transition to a softer state in 1998 (see Nowak, Wilms
\& Dove 1999; Wilms et al. 1999; Belloni et al. 1999;
Nowak, Wilms \& Dove 2002; Corongiu et al. 2003).
In 1999, the source went into quiescence, where it was detected with Chandra
BeppoSAX at low flux levels (Kong et al. 2000; Corbel et al. 2003).

After roughly one year in quiescence, a new outburst started in 2002 
(Smith et al. 2002a,b; Nespoli et al. 2003;
Belloni 2004; Homan et al. 2005) and ended in 2003 (Buxton \& Bailyn 2004a).
This new outburst
was followed in detail through timing and color analysis (Belloni et al. 2005),
which made it the prototype for the HID evolution and the definitions
of the HIMS/SIMS. In particular, a very clear HIMS-SIMS transition was observed
on 2003 May 17, with fast changes in the variability properties, but almost
no variations in the 3-20 keV energy spectrum (Homan et al. 2005).
A relativistically broadened iron emission line has been detected in the
X-ray spectrum of GX 339--4, indicating the presence of a non-zero angular
momentum in the black hole (Miller et al. 2004a,b).
A comparative spectral analysis of all existing RXTE data from GX 339--4
is presented by Zdziarski et al. (2004).

GX 339--4 was also the first BHC showing radio/X-ray and near-IR/X-ray correlations in the LS
(Hannikainen et al. 1998; Corbel et al. 2003; Markoff et al. 2003;  Homan et al. 2005).
Radio observations during the 1999 HS showed clear evidence of a
strong decrease of core radio emission during this state (Fender et al. 1999).
During the 2002/2003, near to the transition to
the VHS (see Smith et al. 2002c), a bright radio flare was observed
(Fender et al. 2002), which led to the formation of a large-scale
relativistic jet (Gallo et al. 2004). Fender, Belloni \& Gallo (2004)
associated this flare and subsequent ejections with the crossing of the
jet line, i.e. the HIMS-SIMS transition reported by Nespoli et al. (2003).

A high mass function (5.8$\pm$0.5 $M_\odot$) has been measured
for the system (Hynes et al. 2003), indicating strong dynamical
evidence for the black-hole nature of the compact object. The distance to
GX 339--4 is not well known, with a lower limit of $\sim$6 kpc (Hynes et al. 2004; see
also Zdziarski et al. 1998).

It is clear that GX 339--4 is an extremely important source for our
understanding of the accretion and ejection properties of stellar-mass
black holes. In February 2004, after having returned to quiescence, GX 339--4 
started a new outburst (Buxton et al. 2004, Smith et al. 2004,
Belloni et al. 2004,  Kuulkers  et al. 2004, Israel et al. 2004). 

\section{Observations}

In August 2004, X-ray and infrared observations of GX 339--4, compared to the
previous 2002/2003 outburst, indicated that the source had entered the HIMS
and was softening (Buxton \& Bailyn 2004b, Homan 2004). 
This suggested an oncoming transition to the SIMS (see Smith \& Bushart 2004) and
prompted us to trigger our RXTE/INTEGRAL campaign. Both RXTE and INTEGRAL
observed the source on 2004 August 14-16. We report here the results of the
joint analysis of RXTE/PCA, RXTE/HEXTE and INTEGRAL/IBIS data. 

\subsection{RossiXTE}
Throughout  our campaign,
RXTE observed GX 339--4 for seven satellite orbits. We also analyzed four
observations from the RossiXTE public archive, two were done before and the two after
our observations. In addition, we extracted background-subtracted PCU2 count rates
from a number of public observations in the time interval MJD 53195-53240.
The log of the observations is shown in Tab. \ref{log}.
We extracted PCA and HEXTE background-corrected energy spectra from each of
the  intervals in Tab. \ref{log} using the standard RXTE software contained in
the package HEASOFT 6.0, following the standard procedures.
For HEXTE, we limited our analysis to cluster A.
To account for residual uncertainties in the calibration, we added a
systematic error of 0.6\% to the PCA spectra. When necessary (see below), we
added spectra using the standard tool {\tt mathpha}.
For the production of the outburst light curve and the corresponding HID, we
accumulated background corrected PCU2 rates in the channel bands $A$=6-48
(2.5-20.2 keV), $B$=6-14 (2.5-6.1 keV) and $C$=23-44 (9.4-18.5 keV), defining
the hardness as  $H=C/B$ (see Homan \& Belloni 2005).
For the timing analysis of the PCA data, for each of the observation intervals
we produced power spectra from 16s stretches accumulated in two channel bands:
0-35 (2-15 keV) and 36-96 (15-40 keV) with a time resolution of 1/128 and 1/2048
seconds. This resulted in eleven spectrograms for each
energy band (see Nespoli et al. 2003). The power spectra were normalised
according to Leahy et al. (1983) and converted to squared fractional rms
(Belloni \& Hasinger 1990).
For different time selections (see below), we averaged the power spectra and
contribution due to Poissonian statistics (see Zhang et al. 1995).
The timing analysis was performed with custom software.

\subsection{INTEGRAL}

As part of the open time program (AO2), INTEGRAL (Winkler et
al. 2003) observed GX 339--4
from Aug 14 21:14 UT to Aug 16 05:29 UT. This 102 ks observation was performed
with a 5x5 dither pattern for a total of 47 pointings, namely ``science
windows'' (SCW), ranging in duration from 1967 s to 4476 s.
In each SCW the target is located in a different part of the instruments field
of view. Thanks to the large IBIS field of view (Ubertini et al. 2003), GX
339--4 was always in its fully coded field of view
(9$^\circ$$\times$9$^\circ$) where the nominal sensitivity of coded mask
instruments remains constant. On the contrary, due to the smaller field of
view and the dithering stategy, the JEM-X coverage of the
sky is considerably smaller (Lund et al. 2003).

In this paper we concentrate on the analysis of data collected with 
ISGRI (Lebrun et al. 2003), the low energy detector layer of IBIS.
The IBIS/ISGRI scientific analysis has been performed usig the
\texttt{ibis\_science\_analysis} main script, included in the latest release
of the INTEGRAL off-line analysis software, OSA 5.0. 
The ISGRI light curves with a binning time of the SCW duration were obtained
extracting the count rate from images. The spectra have been extracted SCW by
SCW with the \texttt{ii\_spectra\_extract} script in 16 logarithmic bins spanning
from 13 keV to 1 MeV.
The response matrices (RMF and ARF) used for spectral fitting 
have been delivered with OSA 5.0 distribution.  

\begin{figure}
\begin{center}
\includegraphics[angle=0,width=9cm]{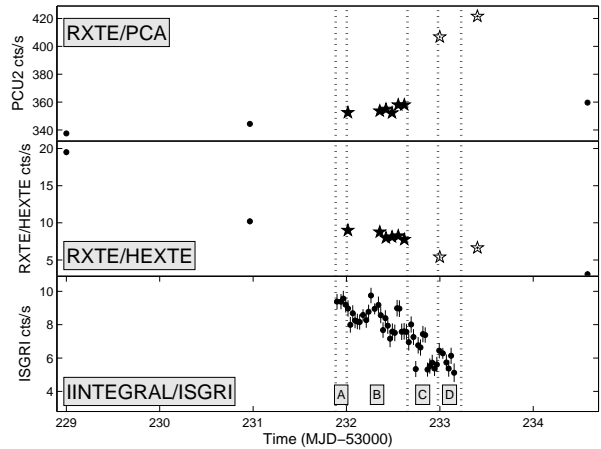}
\end{center}
\caption{RossiXTE and INTEGRAL light curves of GX 339--4 during our campaign, plus two RossiXTE observations before and two after our points. 
The energy ranges are 2.9-20.6 keV (PCA), 20-40 keV (HEXTE and ISGRI). The dotted lines separate the four spectral intervals used for spectral analysis (see text).
\new{Filled stars mark the observation overlapping with INTEGRAL before the transition.
Empty stars indicates the points corresponding to Obs. \#9 and \#10. }
}
\label{licu}
\end{figure}

\section{Results}

In Fig. \ref{licu}, we plot the light curves of the three instruments
(2.9-20.6 keV for PCA, and 20-40 keV for HEXTE and ISGRI). The high-energy
curves (HEXTE and ISGRI) show a decrease by a factor of two between MJD 53229
and 53231, followed by another drop by a factor of two during the MJD 53232
observations.
After a small recovery, the HEXTE flux drops further.
In the PCA band, however, the evolution is different. After a moderate increase
until MJD 53232 (Obs. \#8), a much faster brightening is observed at the
beginning of MJD 53233 (Obs. \#9),  by a factor of $\sim$15\%.
Notice that the sparseness of the PCA points  does not allow us to follow the
evolution between Obs. \#8 and Obs. \#9. The observed evolution suggests that
a transition took place between them.

\begin{figure}
\begin{center}
\includegraphics[angle=0,width=9cm]{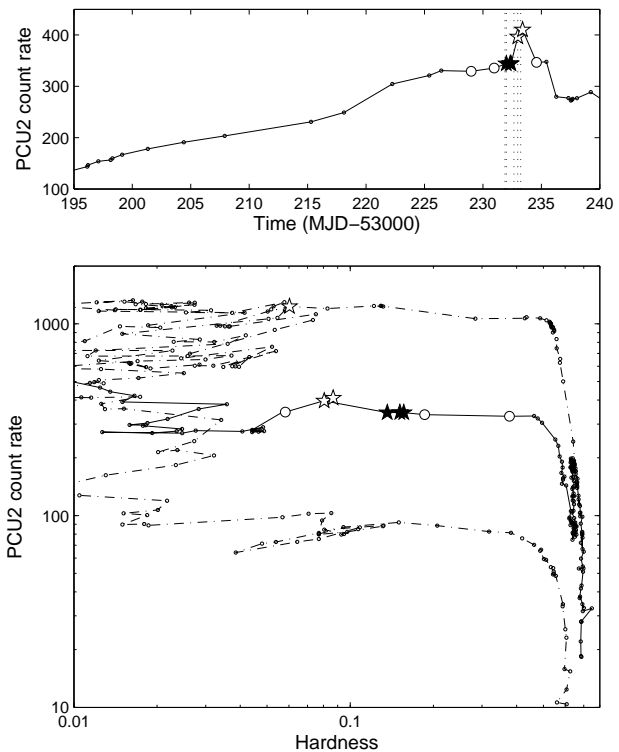}
\end{center}
\caption{Top panel: PCU2 2.9-20.6 keV net light curve of GX 339--4 from the
  public RossiXTE archive until a few days after the observations described
  here. The symbols follow the same convention as in Fig. \ref{licu}.
Bottom panel: Hardness-Intensity diagram from the 2004 outburst (continuous line, outburst not
complete here)
and from the 2002/2003 outburst (dot-dashed line). The time sequence for both curves
  starts at the bottom right. For the 2002/2003 curve, the empty star marks the first SIMS
  point (Nespoli et al. 2003; Belloni et al. 2005).
}
\label{outburst}
\end{figure}

In order to understand our observations in terms of the global evolution of
the outburst, we accumulated from the public RossiXTE archive a light curve
and a HID with the parameters described in the previous section, chosen to be
compatible with the definitions used by Homan \& Belloni (2005). They can be
seen in Fig. \ref{outburst}. Notice that the currently available public data only cover part of the outburst and for this reason the left and bottom branches for this outburst are missing.
It is evident from Fig. \ref{outburst} that the evolution is very similar to that of the previous 2002/2003 outburst (see also Belloni et al. 2005): a monotonic increase of count rate at a rather high
color (corresponding to the LS), a horizontal branch with the source softening
at a nearly constant count rate (the HIMS), further softening with a transition to
the SIMS (see below), 
and further observations at very low hardness,
corresponding to the HS.
Notice how the change in hardness between Obs. \#3-8 and Obs. \#9  is rather
large and takes place on a time scale of a few hours. However, a striking difference
between this outburst and the previous ons is the count rate level of the horizontal/transitional branch,
which here is a factor of $\sim$3.5 lower. 

\begin{table}
\begin{tabular}{lcccc}
\hline
\textbf{Obs.} & \textbf{Type} & \textbf{Start (UT)} & \textbf{End (UT)} &  \textbf{Exp. (s)} \\

\hline
\#1  & P & Aug 11 23:47 & Aug 12 00:05 & 1080\\
\#2  & P & Aug 13 23:01 & Aug 13 23:17 &   960\\
\#3  & C & Aug 15 00:11 & Aug 15 00:33 & 1270\\
\#4  & C & Aug 15 08:05 & Aug 15 09:03 & 3460\\
\#5  & C & Aug 15 09:41 & Aug 15 10:38 & 3420\\
\#6  & C & Aug 15 11:14 & Aug 15 12:12 & 3480\\
\#7  & C & Aug 15 12:49 & Aug 15 13:47 & 3480\\
\#8  & C & Aug 15 14:23 & Aug 15 15:21 & 3480\\
\#9  & C & Aug 15 23:49 & Aug 16 00:09 & 1170\\
\#10 & P & Aug 16 09:21 & Aug 16 10:14 & 3180\\
\#11 & P & Aug 17 13:58 & Aug 17 14:13 &  900\\ 

\hline
\end{tabular}
\caption[]{\footnotesize 
Log of the RossiXTE observations analyzed in this work. `C' indicates observations from our
campaign, `P' observations from the public archive. 2004 August 11 00:00UT corresponds to MJD
53228.
}
\label{log}
\end{table}

\subsection{Timing analysis}

\begin{figure}
\begin{center}
\includegraphics[angle=0,width=9.0cm]{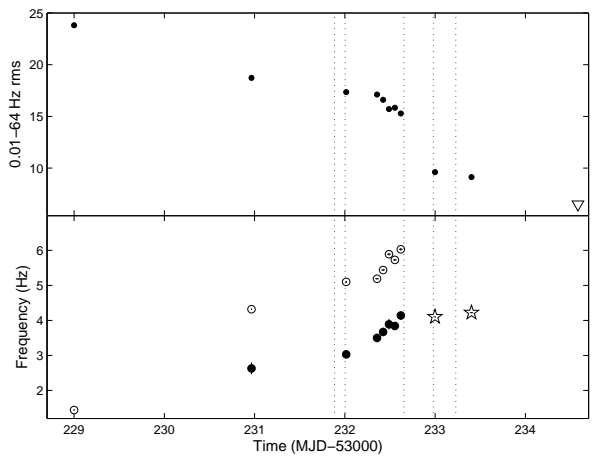}
\end{center}
\caption{Time evolution of selected timing parameters. Top panel: integrated
  0.1-64 Hz fractional rms. Bottom panel: characteristic frequency of the type-C QPO
  (open circles), of the $\sim$4 Hz broad component in Obs. \#3-\#8 (filled
  circle, see Fig. \ref{pds}), and of the type-B QPO in Obs. \#9
  (stars). The inverted triangle represents a 3$\sigma$ upper limit, error
  bars are $1\sigma$.
  The vertical dotted lines correspond to those in Fig. \ref{licu}.
}
\label{timing}
\end{figure}

\begin{figure}
\begin{center}
\includegraphics[angle=0,width=9cm]{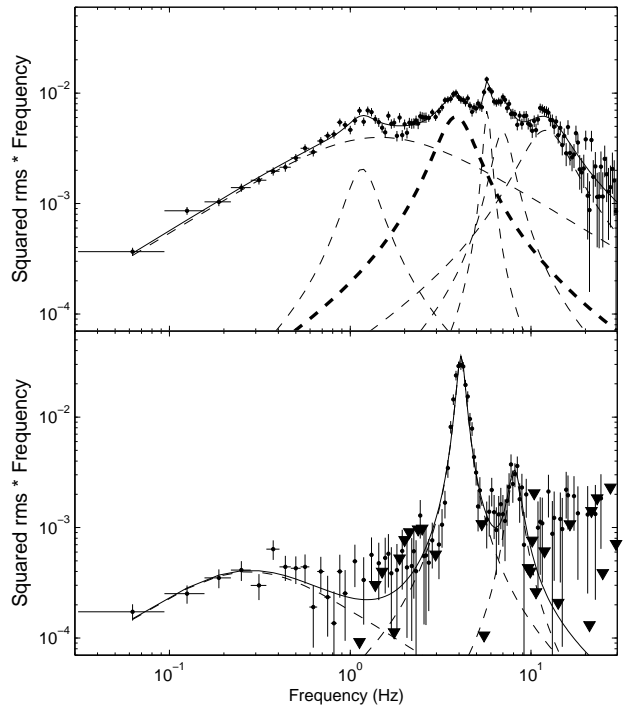}
\end{center}
\caption{Power spectra from PCA data (2-15 keV) of Obs. \#8 (top) and \#9
  (bottom), in $\nu P\nu$ representation. The dashed lines indicate the best
  fit Lorentzian components. The thick dashed line in the upper panel shows
  the component with a frequency similar to that of the main QPO peak in the
  bottom panel. The scales of the two panels are the same.
  \new{The triangles correspond to 3$\sigma$ upper limits.}
}
\label{pds}
\end{figure}

We examined the high-energy power spectra searching for high-frequency
features, but none were found. In the following, by power spectra we mean the
low-energy (2-15 keV) power spectra.
Examining the power spectra, the presence of a transition becomes
evident. 
The power spectrum of Obs. \#1 shows strong band-limited noise, which we
fitted with the sum of two zero-centered Lorentzian components (see Belloni,
Psaltis \& van der Klis 2002). The integrated 0.1-64 Hz fractional rms is 23.8\%. 
The power spectra of Obs. \#2 through \#8 are characterised by an
integrated 0.1-64 Hz fractional rms decreasing from 18.7 to 15.3\% (see top
panel of Fig. \ref{timing}) and complex shape. We fitted it with a combination
of six Lorentzian components. The power spectrum of Obs. \#8, together with
the best-fit Lorentzian models, can be seen in Fig. \ref{pds}. The sharp QPO peak
visible at 6 Hz, appearing together with the band-limited noise component at $\sim$1 Hz,
can be identified with the type-C QPO (called L$_q$ in Belloni et al. 2002,
see Casella et al. 2005) and the corresponding low-frequency
component L$_b$, with characteristic frequencies (see Belloni et al. 2002 for
a definition) separated by a factor of $\sim$5 (see Wijnands \& van der
Klis 1999). The L$_b$ component requires a second narrower Lorentzian for a good
fit. The narrow QPO, as usual, shows a second harmonic peak at twice its
frequency
\new{(its best-fit frequency was consistent with twice that of the L$_q$ component and
was therefore fixed to twice that value)},
and on its high-frequency flank shows an additional component
(L$_h$ in Belloni et al. 2002). There is however a sixth component, with a
characteristic frequency of $\sim$4 Hz, which cannot be obviously identified
(marked with a thick line in Fig. \ref{pds}).
\new{Its frequency is not consistent with being half that of the high-frequency flank
described above, indicating that these two components are not harmonically 
related.}
The evolution of the centroid frequency of the type-C QPO and of this
additional component are shown in the bottom panel of Fig. \ref{timing}: both
of them increase with time, when at the same time the source softens (see
Fig. \ref{licu}), which is a behaviour observed in all black-hole transients
in the HIMS. Indeed, the shape of the power spectrum, with strong band-limited
noise and a type-C QPO, indicate that GX 339--4 was in the HIMS during the
first part of the campaign (Obs. \#2 through \#8), after one observation (\#1)
which did not show a significant QPO and could be classified as LS.
\new{Notice from Fig. 3 (bottom panel) that the frequency of the C-type QPO increases
with time until reaching a value of $\sim$6 Hz, which appears to be  a critical
value for type-B QPOs (see Casella, Belloni \& Stella 2005). It is not clear whether
this value is a coincidence, as in other systems the C-type QPO frequency just before the
transition is higher (see e.g. Casella et al. 2004)}

The power spectrum of Obs. \#9, about nine hours after \#8, appears
completely different (see bottom panel in Fig. \ref{pds}). The noise level is
much lower and can be fitted with one single low-frequency weak Lorentzian,
and only one strong QPO component at 4.10 Hz is present, with an additional
first harmonic peak
\new{(which once again was fixed for the fits to have a centroid frequency twice that of the 
fundamental)}. The integrated fractional rms drops to less than 10\%
(see Fig. \ref{timing}). The peak can be identified with a type-B QPO (see
Wijnands, Homan \& van der Klis 1999; Remillard et al. 2002;
Casella et al. 2005) and this power spectrum, together with the  marked
softening visible in Fig \ref{licu} clearly indicates that the source has
entered the SIMS (Belloni et al. 2005; Homan \& Belloni 2005; Belloni
2005). Notice that the centroid of the type-B QPO (4.10$\pm$0.01 Hz) is
compatible with the characteristic frequency of the additional component in
the power spectrum from Obs. \#8 (4.14$\pm$0.12 Hz, 1$\sigma$ errors), suggesting a possible
identification of the latter.
Obs. \#10 shows a very similar power spectrum. 
The power spectrum of Obs. \#11 shows no significant noise, with a 3$\sigma$
upper limit of 6.5\% to the 0.1-64 Hz fractional rms, suggesting that the
source entered the HS.

\subsection{Spectral analysis}

Given the sparseness of the \#3-\#9 RXTE observations and the evolution of the
flux in the three instruments, we divided the data into four intervals, whose
time limits are shown in Tab. \ref{intervals}. The intervals were defined in
this way: from the start of the INTEGRAL data until Obs. \#3 (interval A),
covering Obs. \#3-\#8 (B), between Obs. \#8 and \#9 (C), and from Obs. \#9 to
the end of the INTEGRAL data (D).
With this definition, intervals B and D contained both RossiXTE and INTEGRAL
data, while intervals A and C were only covered by INTEGRAL.
\new{
We included Obs. \#3 in interval B because including it in interval A resulted in 
a best fit completely compatible with that for interval B. Therefore, we decided
for this subdivision which maximizes the statistics for interval B.
}
In order to obtain a good broad-band spectrum of GX 339--4, we had to limit our
analysis to intervals B (corresponding to the HIMS) and D (SIMS). The
remaining intervals were covered, although sparsely, by the JEM-X1 instrument
on board INTEGRAL. However, to date it was not possible to extract good spectra
from that instrument, and the light curve from the full observation was not compatible
with the PCA one, after restriction to the same energy band.
Therefore, in this work we limit out analysis to PCA+HEXTE+ISGRI.
For the spectral fits, we used the {\tt XSPEC} package version 11.2.3l.

\begin{table}
\begin{tabular}{lcc}
\hline
\textbf{Interval} & \textbf{Start (UT)} & \textbf{End (UT)} \\
\hline

A & Aug 14 21:14  & Aug 15 00:04\\
B & Aug 15 00:05 & Aug 15 15:40\\
C & Aug 15 15:42 & Aug 15 23:30\\
D & Aug 15 23:32 & Aug 16 05:29\\

\hline
\end{tabular}
\caption[]{\footnotesize 
The four time intervals used for spectral extraction of both RossiXTE and
INTEGRAL data.  2004 August 11 00:00 UT corresponds to MJD
53228.
}
\label{intervals}
\end{table}

\begin{figure}
\begin{center}
\includegraphics[angle=0,width=9cm]{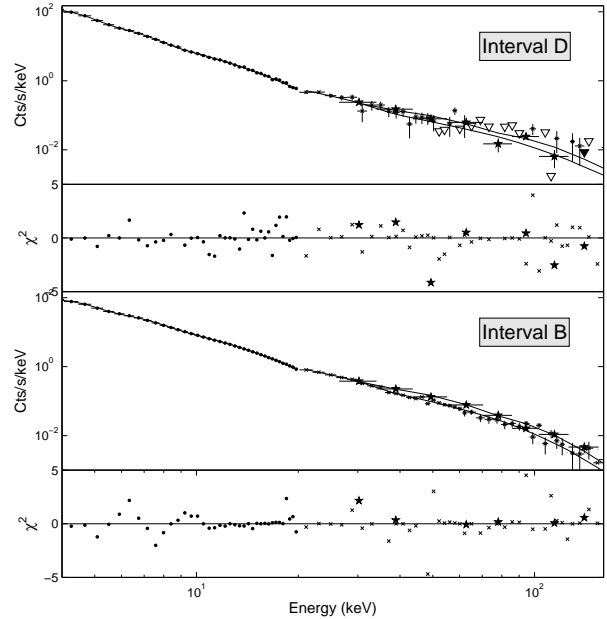}
\end{center}
\caption{Spectra (PCA: filled circles, HEXTE: crosses; ISGRI: stars) from intervals B and D, with best fit model and residuals.}
\label{spectra}
\end{figure}

\begin{figure}
\begin{center}
\includegraphics[angle=0,width=9cm]{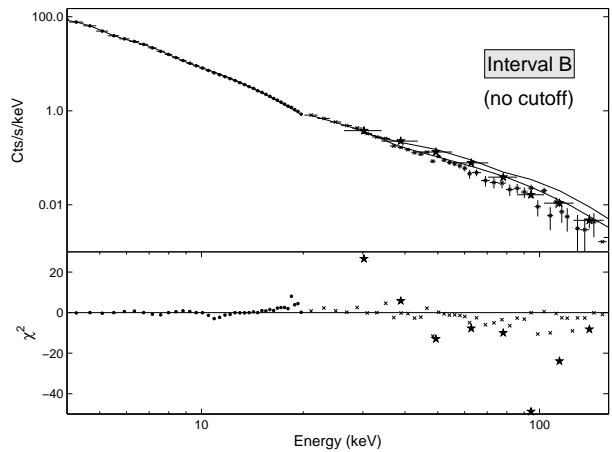}
\end{center}
\caption{Spectra (PCA: filled circles, HEXTE: crosses; ISGRI: stars) from interval B fitted with a power law without a high-energy cutoff  (with best fit model and residuals). The presence of large residuals at
high energies is evident.}
\label{nocutoff}
\end{figure}

We fitted the two spectra (interval B and D) with a simple model consisting of
a disc-blackbody, a power law and an emission line fixed at 6.4 keV, all
modified by interstellar absorption. Since the estimated absorption for GX
339--4 is rather low and we did not have low-energy coverage, we fixed it to
5$\times 10^{21}$ cm$^{-2}$ (M\'endez \& van der Klis 1997; Kong et al. 2000). 
In order to account for remaining cross-calibration problems, we included 
multiplicative constants for the HEXTE and ISGRI spectra with respect to the
PCA. The best fit values we obtained were 0.9 and 1.1 respectively.
The fits with this model gave different results for the two spectra. 
The spectrum of interval D was fitted well by this model, with a reduced
$\chi^2$ of 1.10 for 85 degrees of freedom (see top panel of
Fig. \ref{spectra} and Table \ref{spectral_table}).
For interval B, no satisfactory fit could be obtained, with a minimum reduced
$\chi^2$ of 3.65 for 85 degrees of freedom. The residuals suggested the
inclusion of a high-energy cutoff (see Fig. \ref{nocutoff}). 
With this addition, the reduced $\chi^2$
became 1.07 (for 84 degrees of freedom), a clear improvement,
with a best-fit high-energy cutoff of $72_{-6}^{+10}$ keV. The best-fit
parameters are shown in Tab  \ref{spectral_table}. The main spectral changes
were:  slight softening of the disc-blackbody component, with increased estimated
inner radius by a factor of 1.6$\pm$0.3, steepening of the power-law component, and
disappearance of the $\sim$70  keV high-energy cutoff. 
In order to set an upper limit to the energy of a possible high-energy cutoff
also in the spectrum from interval D, we added such a cutoff and determined the lower
limit to its best fit value. We obtained a 90\% lower limit of 84 keV.
Notice that, although our model of a simple cutoff power law fits the data of interval B
satisfactorily, the presence of an additional Compton reflection component with
a variable reflection fraction might qualitatively yield the same effect observed here.
However, such a complication to the model is not required by our data.
\new{The equivalent width of the 6.4 keV line is 660 eV and 570 eV for interval B and
D respectively. Unlike in the case of the sharp transition of 1E 1740.7--2942 observed in 
May 2001 (Smith, Heindl \& Swank 2002), the photon fluxes show variations roughly consistent
with those of the energy flux.}

\section{Discussion}

From our results, it is clear that we observed a transition from the HIMS to
the SIMS in GX 339--4 on 2004 August 15. The power spectra before and after the
transition show marked differences: from strong band-limited noise and type-C
QPO to much weaker noise and type-B QPO. 
Thanks to the INTEGRAL coverage of more than one consecutive day,
and the simultaneous but more sparse RossiXTE data, we could for the first
time follow the transition at high energies and examine in detail the
broad-band spectrum before and after the transition.
As usual, the transition took place along a nearly horizontal branch
in the HID, although at a different (harder) color position than in 2002 (see Belloni et
al. 2005 and Fig. \ref{outburst}). 

These changes in timing 
properties were also observed in the same
source during the previous outburst in 2002 (Nespoli et al. 2003; Belloni et
al. 2005) and is seen in many black-hole transients (Homan \& Belloni 2005;
Belloni 2005). 
Comparing the properties of this transition and those of the analogous
transition in the 2002/2003 outburst, we can see that, although the shape of
the HID is extremely similar (see Belloni et al. 2005; Homan \& Belloni 2005),
the count rate level of the upper branch is lower by a factor of
$\sim$3.5 (see Fig. \ref{outburst}). 
Also, the color at which the transition takes place is not the
same: in 2002 it was in the range 0.06-0.08 (notice that the color definition of Belloni et al. 2005
is different than the one adopted here: the colors reported here have been thus re-calculated), 
while in our data it's between
0.08 and 0.14 (see Fig. \ref{outburst}). This is reflected by the spectral
parameters (see below). The fact that the HIMS branch, while very similar, is
substantially lower than in 2002 indicates once more that the transition is
not driven by accretion rate only, but another parameter must be at play in
causing it. 
Overall, the evolution through the HID follows similar patterns as all other
bright transients observed by RossiXTE, which underwent state transitions (see
Homan \& Belloni 2005; Belloni 2005).

\begin{table*}
\begin{tabular}{lcccccc}
\hline
Int.&\multicolumn{3}{c}{Disc blackbody} &\multicolumn{3}{c}{(Cutoff-)power law}\\
\hline
 & 
 \textbf{kT [keV]} 
 & \textbf{R$_{in}$ [km]} 
 & \textbf{F$_d$ [erg cm$^{-2}$s$^{-1}$]} 
 & \textbf{$\Gamma$}
 & \textbf{E$_c$ [keV]}
 & \textbf{F$_p$ [erg cm$^{-2}$s$^{-1}$]} \\
\hline

B & 0.90$\pm$0.05   & 13$\pm$2 & 5.7$\pm 3.8\times 10^{-10}$ &1.92$\pm$0.05 & 
72$_{-6}^{+10}$& 2.7$\pm 0.5\times 10^{-9}$\\

D & 0.81$\pm$0.03  & 21$\pm$3 & 7.4$\pm 3.0\times 10^{-10}$&2.31$\pm$0.07 &
         ---                 & 2.2$\pm 0.5\times 10^{-9}$\\

\hline
\end{tabular}
\caption[]{\footnotesize 
Best fit spectral parameters for intervals B and D. For R$_{in}$, the assumed
distance and inclination are 6 kpc and 15$^\circ$ (see Wu et al. 2001). The fluxes are unabsorbed
in the 4-200 keV band. Errors represent 90\% confidence limits.
}
\label{spectral_table}
\end{table*}

The evolution of the power spectra is very similar to that of the previous
outburst, but here we have the advantage of having a good number of RossiXTE
observation intervals just before the transition, so that it was possible to
follow the evolution of the timing components. What we found is that, as
expected, the multiple components in the power spectrum increase their
characteristic frequencies. All these features can be identified with known
components (Belloni et al. 2002), with the exception of a broad bump around  a
few Hz (see Fig. \ref{pds}). The characteristic frequency of this component
increases steadily until it reaches a frequency compatible with that of the
type-B QPO that appears after the transition (see Fig.\ref{timing}). It is therefore possible to
speculate that these components are related, although we would need more
observations of such transitions to asses this identification. 

The comparison of the light curves at low and high energies (see
Fig. \ref{licu}) shows that at high energies ($>$20 keV), where the hard
spectral component dominates, the transition is smooth and takes place over a
longer time scale than at low energies. The HEXTE light curve shows that a
considerable reduction in flux (a factor of $\sim$2) took place in a few days
before the transition, and the ISGRI data show a smooth decrease by the same
factor within one day {\it across} the transition. In other words, the
high-energy curves do not allow to locate the transition. On the other hand the transition is
much more abrupt at low energies (see the hardness changes in
Fig. \ref{outburst} and the timing changes). Although we do not have
complete information on the transition at low energies, due to the limited
coverage, we know from other systems that the switch between type-C and type-B
QPOs can take place over a few seconds \new{(see e.g. Casella et al. 2004)}.
\new{The fact that spectral transitions are slower than timing transitions was already
reported by Kalemci et al. (2004) and can also be seen in the color/timing evolution
of GX 339-4 in its previous outburst (Nespoli et al. 2003, Belloni et al. 2005).}

Particularly important are the joint RossiXTE/INTEGRAL spectral fits. The
amount of information on the high-energy part of the spectrum in different
states of black-hole transients is patchy. In the LS, the broad-band spectrum
is well studied, both with simple phenomenological models and with complex
detailed physical models (see e.g. 
Frontera et al. 2001, Zdziarski et al. 2001, Merloni \& Fabian 2002,
Turolla, Zane \& Titarchuk 2002, Malzac, Merloni \& Fabian 2004, Falcke, K\"ording \&
Markoff 2004).

The main features of the LS spectra
are a rather flat photon index ($\Gamma\sim$1.6) and the clear presence of a
high-energy cutoff around 100 keV.
The high-energy spectrum of the HS is less well known, due to its
softness. Observations with OSSE showed that in states different than the LS,
there is no evidence of a high-energy cutoff up to $\sim$1 MeV (Grove et
al. 1998). However, the integration times needed to accumulate statistically
significant OSSE spectra are of the order of weeks, and these observations
also cover other states such as the HIMS and the SIMS, where the hard
component is stronger and flatter. Therefore, it is difficult to assess the
shape of the high-energy spectra of these states, as transitions can take
place on short time scales. 
Our results constitute one of the first {\it bona-fide} broad-band spectral
determination of the HIMS and the SIMS on such a short time scale. 
The disappearance of the high-energy cutoff in the HEXTE energy band after the
transition to the SIMS was already reported for XTE J1650--500, albeit on a slightly 
longer time scale (Rossi et al. 2005). The HIMS, which can be considered as
an extension of the LS, shortly before the transition to the SIMS, shows a
spectral slope of $\sim$1.9 and a high-energy cutoff around 70 keV. The
spectrum is therefore steeper than in the LS and the high-energy cutoff is at
lower energies. After the transition, in the SIMS, the power-law is
considerably steeper ($\Gamma\sim$2.3), although slightly less than was  observed in the previous
outburst ($\Gamma$=2.44, Nespoli et al. 2003), and there is no evidence of a high-energy
cutoff, with a 90\% lower limit of 84 keV (in 2002 there was
some evidence of a cutoff at $\sim$100--200 keV (Nespoli et al. 2003). 
Although we cannot exclude the presence of a cutoff, our results
show that the cutoff measured in the HIMS either increases significantly in energy
after the transition, or disappears.
As to the 4-200 keV luminosity of the system, adopting a distance of 6 kpc (see
Hynes et al. 2004), the HIMS has a luminosity of 
2.3$\pm 1.5\times 10^{36}\,$erg/s (disc) and 
1.1$\pm 0.2\times 10^{37}\,$erg/s (powerlaw), 
while in the SIMS the disc flux becomes
3.0$\pm 1.2\times 10^{36}\,$erg/s
and the power law flux 9.0$\pm 2.1\times 10^{36}\,$erg/s.
The errors due to the flux reconstruction are still rather large and at 90\% confidence
the disc and power-law luminosities before and after the transition are consistent.
These total disc + power law luminosity are therefore around 1-2\% of
the Eddington luminosity (for 6$M_\odot$).
Notice that these values are considerably lower than those
in Nespoli et al. (2003) for the 2002 transition, as suggested by the difference in 
detected count rate.
Also, here the power-law
component dominates the flux, while in 2002 it did not: this is also reflected
by the difference in hardness at the transition.
In comparison to what was observed in 2002 (Gallo et al. 2004; Fender, Belloni \& Gallo 2004),
we would have expected a radio flare to be triggered by the transition
(Homan 2004), but unfortunately we could not obtain radio coverage.

\section{Conclusions}

The results presented above constitute one of the first determination of the changes of the
broad-band X-ray spectrum of a BHC across the HIMS-SIMS transition, which is crucial
for the development and testing of theoretical models.
Although the transition from the LS to the HS is a process that takes 
days to weeks (see e.g. Belloni et al. 2005), a sharp transition in the properties of 
fast time variability takes place on a much shorter time scales. We showed that this
transition corresponds also to a change in the high-energy properties.

\section*{Acknowledgments}
We thank M. Falanga, A. Bazzano, A. Paizis and I. Kreykenbohm for help in the
INTEGRAL analysis.  This work was partially supported by ASI grants I/R/046/0,
I/R/389/02 and I/R/041/02.

\label{lastpage}


\begin{thebibliography}{99}

\bibitem[]{}
Belloni, T., 2004, in ``The Restless High-Energy Universe'', E. P. J. van den Heuvel, R. A. M. J. Wijers, 
J. M. M. in 't Zand Eds., p337

\bibitem[]{}
Belloni, T., 2005, in Proc. of COSPAR Colloquium ``Spectra and Timing of Compact 
X-Ray Binaries'', Mumbai, India, in press (astro-ph/0507556)

\bibitem[]{}
Belloni, T., Hasinger, G., 1990, A\&A, 230, 103

\bibitem[]{}
Belloni, T, Psaltis, D., van der Klis, M., 2002, ApJ, 572, 392

\bibitem[]{}
Belloni, T., van der Klis, M., Lewin, W.H.G., van Paradijs, J., Dotani, T.,
Mitsuda, K., Miyamoto, S., 1997, A\&A, 322, 857

\bibitem[]{}
Belloni, T., M\'endez, M., van der Klis, M., Lewin, W.H.G., Dieters, S., 1999, 
ApJ, 519, L159

\bibitem[]{}
Belloni, T., Homan, J., Cui, W., Swank, J., 2004, ATel, 236

\bibitem[]{}
Belloni, T., Homan, J., Casella, P., van der Klis, M., Nespoli, E., Lewin, W.H.G., 
Miller, J.M., M\'endez, M., 2005, A\&A, 440, 207

\bibitem[]{}
Brocksopp, C., Fender, R.P., McCollough, M., Pooley, G.G., Rupen, M.P., Hjellming, R.M., 
de la Force, C.J., et al., 2002, MNRAS, 331, 765

\bibitem[]{}
Buxton, M., Bailyn, C., 2004a, ATel, 270

\bibitem[]{}
Buxton, M., Bailyn, C., 2004b, ATel, 316

\bibitem[]{}
Buxton, M., Gallo, E., Fender, R. P., Bailyn, C., 2004, ATel, 230

\bibitem[]{}
Cadolle Bel, M., Sizun, P., Goldwurm, A., Rodr\'\i  guez, J., Laurent, P., Zdziarski, A.A.,
Foschini, L., et al., 2005, A\&A, in press (astro-ph/0509851)

\bibitem[]{}
Casella, P., Belloni, T., Homan, J., Stella, L., 2004, ApJ, 426, 587

\bibitem[]{}
Casella, P., Belloni, T., Stella, L., 2005, ApJ, 629, 403

\bibitem[]{}
Corbel, S., Nowak, M.A., Fender, R.P., Tzioumis, A.K., Markoff, S., 2003, A\&A, 400, 1007

\bibitem[]{}
Corongiu, A., Chiappetti, L., Haardt, F., Treves, A., Colpi, M., Belloni, T., 2003, A\&A, 408, 347

\bibitem[]{}
Dove, J.B., Wilms, J., Maisack, M., Begelman, M.C., 1997, ApJ, 487, 759

\bibitem[]{}
Dove, J.B., Wilms, J., Nowak, M.A., Vaughan, B.A., Begelman, M.C., 1998, MNRAS, 298, 729

\bibitem[]{}
Falcke, H., K\"ording, E., Markoff, S., 2004, A\&A, 414, 895

\bibitem[]{}
Fender, R.P. Belloni, T., Gallo, E., 2004, MNRAS, 355, 1105

\bibitem[]{}
Fender, R.P., Corbel, S., Tzioumis, T., McIntyre, V, Campbell-Wilson, D., Nowak, M.A., Sood, R.,
1999, ApJ, 519, L165

\bibitem[]{}
Fender, R.P., Corbel, S., Tzioumis, T., Tingay, S., Brocksopp, C., Gallo, E., 2002, ATel, 107

\bibitem[]{}
Frontera, F., Palazzi, E., Zdziarski, A.A., Haardt, F., Perola, G.C., Chiappetti, L., 
Cusumano, G., et al., 2001, ApJ, 546, 1027

\bibitem[]{}
Gallo, E., Corbel, S., Fender, R.P., Maccarone, T.J., Tzioumis, A.K., 2004, MNRAS, 347, L52

\bibitem[]{}
George, I., Fabian, A.C., 1991, MNRAS, 249, 352

\bibitem[]{}
Gierli\'nski, M., Zdziarski, A.A., Poutanen, J., Coppi, P.S., Ebisawa, K., Johnson, W.N., 1999,
MNRAS, 309, 496

\bibitem[]{}
Grove, J.E., Johnson, W.N., Kroeger, R.A., McNaron-Brown, K., Skibo, J.G., 
Phlips, B.F., 1998, ApJ, 500, 899

\bibitem[]{}
Hannikainen, D.C., Hunstead, R.W., Campbell-Wilson, D., Sood, R.K., 1998, A\&A, 337, 460

\bibitem[]{}
Haardt, F., Galli, M.R., Treves, A., Chiappetti, L., Dal Fiume, D., Corongiu, A.,
Belloni, T., Frontera, F., Kuulkers, E., Stella, L., 2001, ApJSuppl, 133, 187

\bibitem[]{}
Homan, J., 2004, ATel, 318

\bibitem[]{}
Homan, J., Belloni, T., 2005, in ``From X-ray Binaries to Quasars: Black Hole Accretion on All Mass Scales'', T.J. Maccarone, R.P. Fender, L.C. Ho Eds, Kluwer, Dordrecht, in press 
(astro-ph/0412597)

\bibitem[]{}
Homan, J., Wijnands, R., van der Klis, M., Belloni, T., van Paradijs, J., Klein-Wolt, M., 
Fender, R.P., M\'endez, M., 2001, ApJSuppl, 132, 377

\bibitem[]{}
Homan, J., Buxton, M., Markoff, S., Bailyn, C.D., Naspoli, E., Belloni, T., 2005, ApJ, 624, 259

\bibitem[]{}
Hynes, R. I., Steeghs, D., Casares, J., Charles, P. A., O'Brien, K., 2003, ApJ, 583, L95

\bibitem[]{}
Hynes, R. I., Steeghs, D., Casares, J., Charles, P. A., O'Brien, K., 2004, ApJ, 609, 317

\bibitem[]{}
Ilovaisky, S.A., Chevalier, C., Motch, C., Chiappetti, L., 1986, A\&A, 164, 67

\bibitem[]{}
Israel, G., Covino, S., Kuulkers, E., Zerbi, F.M., Chincarini, G., Rodon\'o, M., Antonelli, L.A.,
2004, ATel, 243

\bibitem[]{}
Kalemci, E., Tomsick, J.A., Rothschild, R.E., Pottschmidt, K., Kaaret, P., 2004, ApJ, 603, 231

\bibitem[]{}
Kong, A.K.H., Kuulkers, E., Charles, P.A., Smale, A.P., 2000, MNRAS, 311, 405

\bibitem[]{}
Kuulkers, E., Bodaghee, A., Foschini, L., Guainazzi, M., Matt, G., Israel, G., Nicastro, F., et al., 
2004, ATel, 240

\bibitem[]{}
Laurent, P. Titarchuk, L., 1999, ApJ, 511, 289

\bibitem[]{}
Leahy, D.A., Darbro, W., Elsner, R.F., Weisskopf, M.C., Kahn, S., Sutherland, P.,G., 
Grindlay, J.E., 1983, ApJ, 266, 160

\bibitem[]{}
Lebrun, F., Leray, J. P., Lavocat, P., Cr\'etolle, J., Arqu\`es, M., Blondel, C., Bonnin, C., et al., 
2003, A\&A, 411, L141

\bibitem[]{}
Lund, N., Budtz-J\o rgensen, Westergaard, N. J., Brandt, S., Rasmussen, I. L., 
Hornstrup, A., Oxborrow, C. A., et al., 2003, A\&A, 411, L231

\bibitem[]{}
Maejima, Y., Makishima, K., Matsuoka, M., Ogawara, Y., Oda, M., Tawara, Y., Doi, K., 1984,
ApJ, 285, 712

\bibitem[]{}
Markoff, S., Falcke, H., Fender, R.P., 2001, A\&A, 372, L25

\bibitem[]{}
Markoff, S., Nowak, M.A., Wilms, J., 2005, ApJ, in press (astro-ph/0509028)

\bibitem[]{}
Malzac, J., Merloni, A., Fabian, A.C., 2004, MNRAS, 351, 253

\bibitem[]{}
Malzac, J., Petrucci, P.O., Jourdain, E., Cadolle Bel, M., Sizun, P., Pooley, G., Cabanac, C., 
et al., 2005, A\&A, submitted

\bibitem[]{}
Markoff, S.. Nowak, M.A., Corbel, S., Fender, R.P., Falcke, H., 2003, A\&A, 397, 645

\bibitem[]{}
McClintock, J.E., Remillard, R.A., 2005, in ``Compact stellar X-ray sources'', W.H.G. Lewin \&
M. van der Klis Eds., Cambridge Univ. Press, Cambridge, in press (astro-ph/0306213)

\bibitem[]{}
M\'endez, M., van der Klis, M., 1997, ApJ, 479, 926

\bibitem[]{}
Merloni, A., Fabian, A.C., 2002, MNRAS, 332, 165

\bibitem[]{}
Miller, J.M., Fabian, A.C., Wijnands, W., Remillard, R.A., Wojodowski, P., Schulz, N.S.,
Di Matteo, T., Marshall, H.L., Canizares, C.R., Pooley, D., Lewin, W.H.G., 2002,
ApJ, 578, 348

\bibitem[]{}
Miller, J.M., Fabian, A.C., Reynolds, C.S, Nowak, M.A., Homan, J., Freyberg, M.J.,
Ehle, M., Belloni, T., Wijnands, R., van der Klis, M., Charles, P.A., Lewin, W.H.G., 2004a,
ApJ, 606, L131

\bibitem[]{}
Miller, J. M., Raymond, J., Fabian, A. C., Homan, J., Nowak, M.A., Wijnands, R., 
van der Klis, M., et al., 2004b, ApJ, 601, 450

\bibitem[]{}
Miyamoto, S., Kimura, K., Kitamoto, S., Dotani, T., Ebisawa, K., 1991, ApJ, 383, 784

\bibitem[]{}
Nespoli, E., Belloni, T., Homan, J., Miller, J.M., Lewin, W.H.G., M\'endez, M., van der Klis, M.,
2003, A\&A, 412, 235

\bibitem[]{}
Nowak, M.A., Wilms, J., Dove, J.B., 1999, ApJ, 517, 355

\bibitem[]{}
Nowak, M.A., Wilms, J., Dove, J.B., 2002, MNRAS, 332, 856

\bibitem[]{}
Nowak, M.A., Wilms, J., Heindl, W.A., Pottschmidt, K., Dove, J.B., Begelman, M.C., 2001
MNRAS, 320, 316

\bibitem[]{}
Pottschmidt, K., Wilms, J., Nowak, M.A., Pooley, G.G., Gleissner, T., Heindl, W.A., Smith, D.M.,
2003, A\&A, 407, 1039

\bibitem[]{}
Remillard, R.A., Sobczak, G.J., Muno, M.P., McClintock, J.E., 2002, ApJ, 564, 962

\bibitem[]{}
Reynolds, C.S, Nowak, M.A., 2003, PhR, 377, 389

\bibitem[]{}
Rodr\'\i guez, J., Corbel, S., Hannikainen, D.C., Belloni, T., Paizis, A., Vilhu, O., 2004, 
ApJ, 615, 416

\bibitem[]{}
Rossi, S., Homan, J., Miller, J.M., Belloni, T., 2005, MNRAS, 360, 763

\bibitem[]{}
Smith, D.M., Bushart, S.K., 2004, ATel, 322

\bibitem[]{}
Smith, D.M., Heindl, W.A., Swank, J.H., 2002, ApJ, 569, 362

\bibitem[]{}
Smith, D. M., Swank, J. H., Heindl, W. A., Remillard, R. A., 2002a, ATel, 85

\bibitem[]{}
Smith, D. M., Belloni, T., Kalemci, E., et al., 2002b, IAUC, 7912

\bibitem[]{}
Smith, D. M., Belloni, T., Heindl, W. A., et al., 2002c, ATel, 95

\bibitem[]{}
Smith, D. M., Heindl, W. A., Swank, J. H., Wilms, J., Pottschmidt, K., 2004a, ATel, 231

\bibitem[]{}
Sunyaev, R.A., Tr\"umper, J., 1979, Nature, 279, 506

\bibitem[]{}
Tananbaum, H., Gursky, H., Kellogg, E., Giacconi A., Jones, C., 1971, ApJ, 177, L51

\bibitem[]{}
Turolla, R., Zane, S., Titarchuk, L., 2002, ApJ, 576, 349

\bibitem[]{}
Ubertini, P., Lebrun, F., Di Cocco, G., Bazzano, A., Bird, A.J., Broenstad, K., Goldwurm, A., 
et al., 2003, A\&A, 411, L131

\bibitem[]{}
van der Klis, M., in ``X-ray Binaries'', W.H.G. Lewin, J. van Paradijs \& E.P.H. van den Heuvel Eds., 
Cambridge Univ. Press, Cambridge, p252

\bibitem[]{}
van der Klis, M., 2005, in ``Compact stellar X-ray sources'', W.H.G. Lewin \& M. van der Klis Eds.,
Cambridge Univ. Press, Cambridge, in press (astro-ph/0410551)

\bibitem[]{}
Wijnands, R., van der Klis, M., 1999, ApJ, 514, 939

\bibitem[]{}
Wijnands, R., Homan, J., van der Klis, M., 1999, ApJ, 526, L33

\bibitem[]{}
Wilms, J., Nowak, M.A., Dove, J.B.,  Fender, R.P., Di Matteo, T., 1999, ApJ, 522, 460

\bibitem[]{}
Wilms, J., Nowak, M.A., Pottschmidt, K., Heindl, W.A., Dove, J.B., Begelman, M.C., 2001,
MNRAS, 320, 327

\bibitem[]{}
Winkler, C., Courvoisier, T.J.-L., Di Cocco, G., Gehrels, N., Gim\'enez, A., Grebenev, S., 
Hermsen, W., et al., 2003, A\&A, 411, L1

\bibitem[]{}
Wu, K., Soria, R., Hunstead, R.W., Johnston, H.M., 2001, MNRAS, 320, 177

\bibitem[]{}
Zdziarski, A.A., Poutanen, J., Miko\l ajewska, J.,  Gierli\'nski, M., Ebisawa, K., Johnson, W.N.,
1998, MNRAS, 301, 435

\bibitem[]{}
Zdziarski, A.A., Grove, J.E., Poutanen, J., Rao, A.R., Vadawale, V., 2001,
ApJ, 554, L45

\bibitem[]{}
Zdziarski, A.A., Gierli\'nski, M., Miko\l ajewska, J., Wardzi\'nski, G., Smith, D.M., Harmon, B.A., 
Kitamoto, S., 2004, MNRAS, 351, 791

\bibitem[]{}
Zhang, W., Jahoda, K., Swank, J.H., Morgan, E.H., Giles, A.B., 1995, ApJ, 449, 930

\end{thebibliography}
\end{document}